\documentclass[times[12pt,twocolumn]{article}
\newcommand{\malware}{malware}
\newcommand{\conficker}{conficker}
\newcommand{\tor}{Tor}
\usepackage{graphicx}
\usepackage{url}
\title{Torinj : Automated Exploitation Malware Targeting Tor Users}
\author{G{\'e}rard Wagener \\ quuxlabs \\ gerard.wagener@quuxlabs.com \and Alexandre Dulaunoy \\ quuxlabs \\ alexandre.dulaunoy@quuxlabs.com\\
\and Radu State \\ University of Luxembourg \\ radu.state@uni.lu}
\begin{document}
\date{24 May 2009}
\maketitle
\abstract{
We propose in this paper
a new propagation vector for malicious software by abusing the \tor\ network.
\tor\ is particularly relevant, since operating a \tor\ exit node is easy and 
involves low costs compared to attack institutional or ISP networks. 
After presenting the \tor\ network from an attacker perspective, we describe 
an automated exploitation \malware\ which is operated on a \tor\ exit node 
targeting to infect web browsers. Our experiments show that the current deployed
\tor\ network, provides a large amount of potential victims.
}
\section{Introduction}
The ubiquitous computation and network infrastructure, currently deployed, is 
exposed to numerous risks. Recently the \conficker\ worm, a self-spreading 
malicious software (\malware) infected millions of machines \cite{conficker}. 
Malware often uses multiple attack vectors. According to the authors of 
\cite{conficker} the \conficker\ worm can also propagate via network shares and USB 
sticks. Moreover some evil web pages infect visitors with malicious software 
\cite{provos08a}. Some users believe that \tor\ \cite{tor}, an anonymous 
communication service, can help to mitigate against privacy and 
confidentiality attacks. \tor\ can be summarized as 
overlay network aiming to hide one's identity which is formally proved \cite{proof}.
According to Eric Cronin  \cite{cronin06} eavesdropping is a difficult task due to 
the fact that packets could be misinterpreted. An additional problem for attackers 
is to wiretap at a strategic point where multiple hosts can be sniffed. 
However, \tor\ simplifies traffic eavesdropping for an attacker. An attacker simply 
needs to install \tor\ exit nodes and participate in the \tor\ network.
Another advantage for an attacker to use \tor, is its proven anonymity which is 
tempting to create a stealthy and anonymous command and control center for 
controlling the eavesdropping and the infection of machines.

In this paper, we propose a novel propagation mechanism of malicious 
software via \tor\ and the contributions of this paper are
\begin{itemize}
    \item an estimation of the vulnerable browsers aiming to tune the web 
          browser infection.
    \item a mechanism to enforce interactions with the web browsers aiming 
          to distribute malicious payloads.
\end{itemize}

The remaining paper is organized as follows:

Section \ref{rw} describes related work and focus on 
potential attacks on \tor. 
An attacker incentive model is presented in section \ref{tm} which  motivates 
the design and implementation of an automatic exploitation \malware\  using the 
Tor network, shown in section \ref{poc}. Section \ref{ccl} concludes the article 
and announces future work activities.  

\section{Related work}\label{rw}
The main purpose of \tor\ is to provide anonymous communication services. This is 
achieved by setting up an overlay network composed of entry guard nodes, 
relay nodes and exit nodes. A client that wants to use the \tor\ network connects 
to entry guard and then establish a circuit towards the exit nodes. In this circuit 
each node only knows its predecessor \cite{tor}. Profiling attacks on encrypted
web proxy traffic were already studied by analyzing the exchanged number of 
bytes \cite{safeweb}. McCoy et al. studied \tor\ traffic \cite{maccoy08}. 
They captured
traffic at entry guards and exit nodes. Thus, they were able to study some clear
text protocols like HTTP and telnet. The purpose of their study was to gain some
insights about the \tor\ usage. In their study they could establish the number of 
different users passing through their entry guard, because they could see where 
they are 
coming from. However, when analyzing traffic from an exit node the traffic is 
already anonymised  which makes it hard to distinguish users.

From that paper can be concluded that the most used 
protocol is HTTP. A threat model for the \tor\ network was proposed by
\cite{tor} and \cite{maccoy08}. An attacker can intercept some fraction of the
traffic. She also can generate, modify, delay some traffic and can compromise
a fraction of the \tor\ nodes. Roger Dingledine et al. described various attacks 
on the different \tor\ nodes \cite{tor} and McCoy et al. even present 
countermeasures to detect
\tor\ exit nodes that are intercepting traffic \cite{maccoy08}. Their major 
assumption is that the
attacker is doing DNS reverse lookups in real time. Furthermore, efforts 
are done to wipe out sensitive information like user agents, cookies from HTTP 
requests. Privoxy \cite{privoxy} is a local proxy implementation that hides some 
sensitive information. An experiment, performed by Dan Egerstad \cite{torhack}, 
showed that a lot of \tor\ users transmit sensitive information, like account 
names, user names and passwords through the \tor\ network without an end to end 
encryption. Security improvements in the Tor network are described by Mike Perry
\cite{tor-sec}
and especially in the area of application attacks at the exit nodes. One of the
proposed improvement is to carefully distribute Tor exit nodes usage to use 
disjoint IP networks. Mike Perry also announced to compute checksums of carefully 
selected  web pages in order to detect injection attacks. 

\section{Attacker Incentive Model}\label{tm}
As discussed in section \ref{rw}, attackers can easily eavesdrop traffic on a \tor\ 
exit node. In this paper
we go a step further and propose an automated exploitation \malware\ that is
capable to infect browsers that pass through an exit node. An attacker should
be able to estimate the population of vulnerable browsers and to enforce an
interaction with the browsers.

\subsection{Passive attacks}
Besides the tools 
like Privoxy that try to wipe out  most of this information some users still 
provide browser information like user agents and cookies. Many browsers set this 
string. 
Some browsers 
start this string by setting the browser's name followed with the version. 
Other browsers set the browser family first and then put the browser 
name between brackets. Furthermore, some browsers provide information about 
the underlying operating system and used libraries. This unorganized user agent
naming provides us some insights about the users that are surfing via our exit node.

Furthermore, the Mitre organization hosts the Common Vulnerabilities and 
Exposure Database (CVE) which contains known software vulnerabilities from 1999 
until now, including  browser vulnerabilities. If we observe $n$ browsers,
$V$ of them are vulnerable and for $\bar{V}$ no vulnerability was reported. 
Thus we can compute 
the browser vulnerability ratio $b$ defined in eq. \ref{bvul}. If all 
observed user agents are vulnerable the browser vulnerability ratio becomes 1,
and if no observed user agents are vulnerable b = 0.

\begin{equation}\label{bvul}
    b = \frac{V}{V+\bar{V}}
\end{equation}

\subsection{Active attacks}
As previously described, the browser infection ratio can be computed. User agent
strings can be forged. Tools like Privoxy change user agent strings. Moreover 
proxies or browsers can be configured to not download external objects attached
to a web site, where an attacker can place his infection payload dedicated for 
web browsers. Hence, a feedback from an observed HTTP traffic is desired.

An attacker can tag HTML responses for getting this feedback. Practically, an
attacker can set up a man-in-the middle attack by installing a transparent 
proxy on the \tor\ exit node.

She can introduce $n$ images or other objects in intercepted HTML documents. 
In case a regular browser is parsing these  pages it tries to acquire the 
objects. Normally the URL of the object  is first resolved followed by the 
download of the object. 

We define a tag as an object that is injected in the intercepted HTML traffic and
we propose two tags per HTML response.

\subsubsection{Static tag injection}
A static tag is a fix invisible image that is introduced in HTTP responses. The 
image has a dimension of 1 to 1 pixel and is invisible aiming to not distract 
the user looking at the HTML page. The URL of the image is fix for all users. 
We assume that the DNS cache on the user's machine is working correctly and that 
the lookup of the image domain name is only done once while the user is surfing. 
Thus we can count the number of different users.

\subsubsection{Dynamic tag injection}
The purpose of the dynamic tag is to enforce an interaction for each visited web
page. In case only a static image is used, the image is normally resolved once
and then it is kept in cache for all the next web pages that are visited during 
the life time of the browser. In order to avoid this caching mechanism an attacker
can generate a unique sub domain for each injected dynamic image. Thus the machine
hosting the browser is forced to do a DNS lookup.
An attacker can also observe 
if a user comes back. In this case the user restarted her machine, reloaded her 
browser with a dedicated web page. In that web page, the attacker previously 
injected an image located on a unique sub domain. Hence, if 
the attacker sees more than one hit for a unique generated sub domain, she can
deduce that the same user reappeared.

\subsection{Attacker Information sources}
By intercepting and tagging HTML documents an attacker can explore three 
information sources.

\begin{description}
\item [DNS server]
We assume that an attacker controls a DNS server for generating unique 
sub-domains for each dynamic tag. The attacker can log all the DNS queries 
including source IP addresses that do the DNS queries.

\item [Web proxy]
The tag injection can be done by doing a man-in-the middle attack. An attacker
can compromise an exit node and set up a transparent proxy for inserting 
the tags. From this web proxy the attacker can record all the HTTP header 
information like user agents or cookies.

\item [TCP traffic]
After having compromised an exit node an attacker can also record all the
out going traffic from the exit code. Thus she has access to the full communications
of \tor\ users. The attacker can focus on HTTP responses, especially on the 
mime type of a message, aiming to tune her browser infection. For instance, if
she notices that most HTTP responses are HTML documents, she could inject images 
in the transferred HTML documents. However, if she sees that the most transfered 
documents are PDF files she could launch PDF attacks.
\end{description}

\section{Torinj : An Automated Exploitation Malware } \label{poc}
\begin{figure}
    \begin{center}
        \includegraphics[scale=0.35]{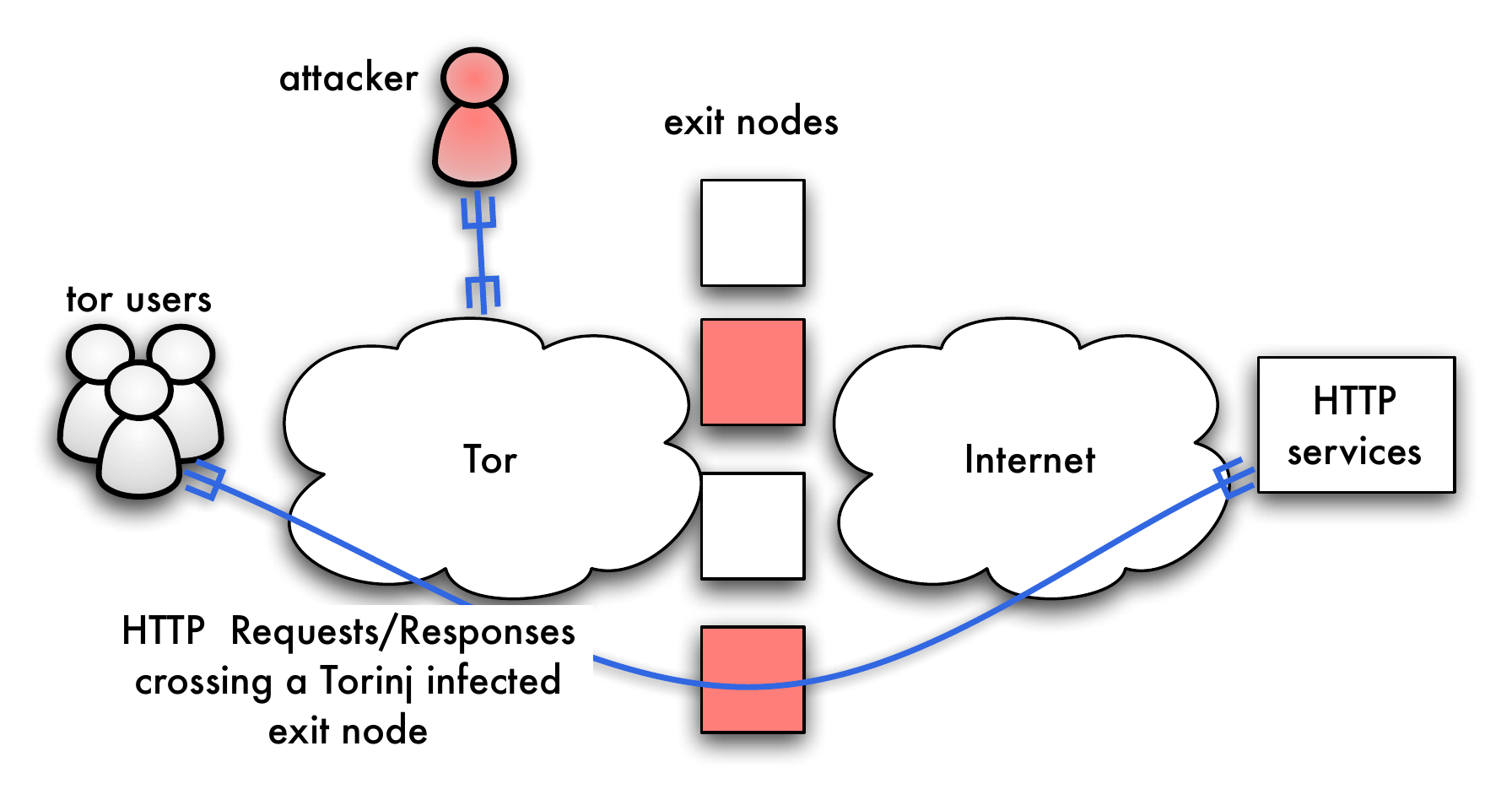}
    \end{center}
\caption{An overview of the Torinj framework}
\label{overview}
\end{figure}

To validate the attacker incentive model we implemented a proof of concept 
\malware\ called Torinj. 
Torinj is composed of three components : an unmodified
Tor client, an embedded intercepting proxy and a hidden C\&C (command and control) 
channel. An overview is shown in figure \ref{overview}.
A standard and unmodified Tor client is integrated with Torinj providing the access
to the Tor network layer. Torinj behaves like any other Tor client and provides
similar services like relay or exit functionalities.
Torinj includes a small HTTP proxy used to intercept and relay HTTP requests. 
Interception and relaying are activated by the attacker using the C\&C channel.
The hidden C\&C channel relies on hidden service protocol \cite{tor-hidserv} 
available in Tor to provide some anonymity \cite{hs-attack06} to the command 
and control interface and its user. The attacker access the C\&C channel of 
each Torinj bot through the Tor network.
Torinj infection is working at the interception level and does not need to lure 
the users to connect to attractive services. 
Torinj infection is done on the unencrypted HTTP requests/responses 
crossing the infected exit node.
The exploitation mechanism of Torinj is composed of two steps:

\begin{description}
    \item [passive attacks,] where the Torinj HTTP proxy is gathering essential 
    information about the HTTP requests (e.g. browser user agent, Internet media 
    type) without altering the requests;

    \item[active attacks,] where Torinj is exploiting the HTTP requests by 
         modifying the responses based on the optimal infection scenario learned 
         by the previous step.
\end{description}

For further technical details we recommend to read our source
code\footnote{\url{http://www.foo.be/torinj/}}

\subsection{Experiment setup}
During this experiment we used three different machines. On the first machine 
$(M_{1})$ we operated an unmodified \tor\ exit node (v0.2.1.14-rc). On the second 
machine we let run BIND \cite{bind}, version 9.4.2, as DNS server and used the tool 
tcpdump \cite{tcpdump} to capture all the DNS queries and responses. On the 
third machine ($M_{3}$) we operated an apache web server \cite{apache}, 
version 2.2.6, hosting the transparent image simulating a malicious payload. 
From a legal and ethical point of view we avoided to inject malicious java
script payloads like XSS-proxy or BeEF \cite{cannings}.

All the machines were synchronized 
with NTP \cite{ntp} in order to have accurate timestamps.
After having started to participate in the \tor\ network, we set up a web proxy
implemented in Perl from CPAN \cite{cpan} (0.23). This proxy was extended 
to inject tags with regular expressions. We used the tool iptables \cite{iptables}
to reroute the traffic, originated from the \tor\ exit node to the Internet, to 
our Perl proxy server.
The DNS server was configured with a wild card that it should associate all 
sub-domains with the IP address of our web server. 
Thus inside the web proxy we can generate dynamic and static tags that always 
point to our web server. As information sources we used tcpdump activated on $M_{1}$
and $M_{2}$, the web server logs, the web proxy logs. The processing was done with 
Perl and sqlite3 \cite{sqlite} and a modified version of tcpick \cite{tcpick}.

\subsection{Passive attacks}
We operated a \tor\ exit node for a period of 28 hours and we passively inspected 
observed HTTP headers. In this experiments we observed similar results to
McCoy et al \cite{maccoy08}. We observed that 96\% of the traffic was HTTP and only 4\% of 
the traffic was end-to-end encrypted with HTTPS.

\begin{figure}
    \begin{center}
        \includegraphics[scale=0.35]{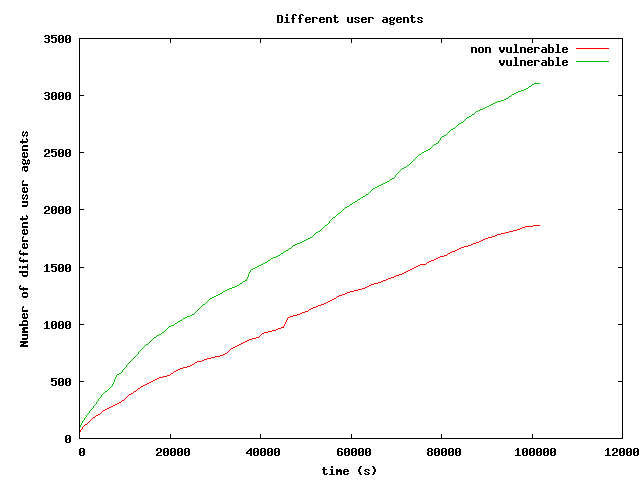}
    \end{center}
\caption{Vulnerable and non vulnerable user agents}
\label{browsers}
\end{figure}

\begin{figure}
    \begin{center}
        \includegraphics[scale=0.35]{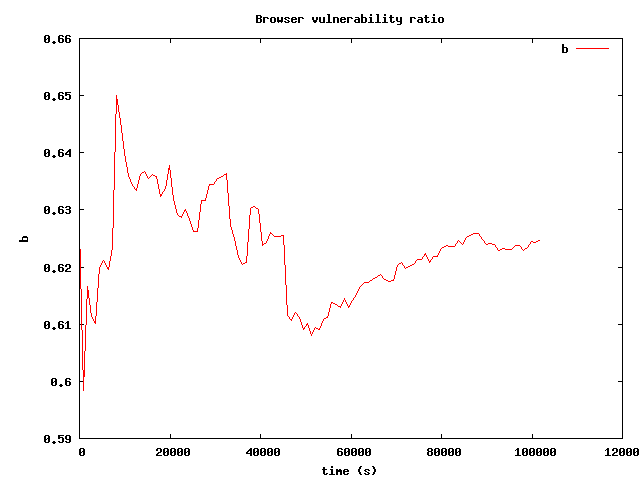}
    \end{center}
\caption{Browser vulnerability ratio}
\label{brinf}
\end{figure}

We have also discovered 4973 different user agent strings which confirms
the non existence of naming convention for user agents. We have found that only 3.2\%
of the HTTP requests did not have a user agent set. We assume that these browsers
are not vulnerable despite they could be vulnerable versions.
Moreover we did an automatic lookup of the user agent in the CVE list. 
Although 1845 user agents did not match an entry in the CVE list (37\% of the
browsers), there may be undisclosed vulnerabilities. If a version is not 
explicitly set for a given browser, we assume that this browser is
not vulnerable (1.3\% of the observed browsers). We found 3106 vulnerable unique 
user agents. Figure \ref{browsers} confirms the fact that there are more 
vulnerable browsers than non vulnerable browsers.  We measured during time intervals 
of 15 minutes the number of vulnerable browsers and non vulnerable browsers.
The number of unique user agent strings is growing (figure \ref{browsers}) because 
most 
user agent strings contain version numbers, with which other browsers they are 
compatible, information about the underlying operating system, patch levels and 
used libraries. Figure \ref{brinf} presents the browser vulnerability ratio, which
is varying around 0.63. That means that on average 63 browsers of 100 are 
vulnerable according the CVE list which shows the  potential of our automated 
exploitation \malware.

\subsection{Active attacks}
For this experiment we set up and operated the automated exploitation \malware\ 
proof of concept for two and a half hours. We have observed 391 different 
user agents that passed through our proxy. The proxy injected 126 static tags and 
688 dynamic tags.

The purpose of the static tag is to count the different users. If a user opens 
her browser, the later resolves the static tag, the static tag is then 
kept in the user's cache
and it should not be resolved again. On our DNS server we observed 126 hits 
for the static image and on our web server we counted 196 hits. Most of the hits
on the web server were done through our proxy. However 80 of the web server hits 
passed trough other \tor\ exit nodes. When the injected tags are downloaded 
through our proxy, our proxy did not tag these responses again. 
We also have counted 391 different user agent strings. This number is 
higher than the number of static tag injection hits and the ratio corresponds to 
32\% which can be explained that only HTML documents were tagged and other 
mime types were directly forwarded without change.

Each HTML document having a body element is intercepted and a unique dynamic tag
is injected and our proxy injected 688 tags.

\subsubsection{Mime type distribution}
In order to get a feedback from a user the injected tag needs to be processed 
by the user agent and the user agent needs to connect back to the attacker. 
This is often the case when HTML documents are processed. Table \ref{mime-dist} 
shows the mime type distribution. Roughly a third of the traffic that goes through
the proxy is composed of HTML documents.

\begin{table}
\begin{tabular}{ll}
Mime type                                    & \%\\
\hline
text/html                                    &33\\
image/jpeg                                   &24\\
image/gif                                    &16\\
image/png                                    &06\\
text/plain                                   &05\\
Content-Type:application/x-javascript        &04\\
Content-Type:text/css                        &03\\
Content-Type:text/javascript                 &03\\
Content-Type:text/xml                        &02\\
Others:                                      &06\\
\end{tabular}
\caption{Mime type distribution}
\label{mime-dist}
\end{table}

\section{Conclusion and future work}\label{ccl}
In this paper, we have described the incentive for an attacker to compromise
Tor exit nodes and designed the Torinj scenario targetting the HTTP protocol.
The experiments further demonstrate the viability of the Torinj prototype 
and the inherent interest for an attacker to compromise Tor exit nodes.
Our experiments 
showed that 63\% of the  browser passing through an exit node are vulnerable 
according the CVE database. Moreover, we showed that interaction
with the browsers can be induced by injecting tags in HTML documents. By 
injecting tags in HTML documents an interaction per web-page can be enforced
which is necessary of malicious payload distribution.
However, additional research efforts are needed to complete this proof of
concept. First of all, the automated exploitation \malware\ should be operated
over a longer period of time and from different IP addresses. We already facilitated 
this work by making our exploitation software freely available under a
GPL license. Furthermore, user agents can be carefully crafted to trick the
exploitation \malware. Therefore other browser finger printing techniques
should be explored. We have only tested the injection in HTML documents and other 
mime types, like PDF, images, movies can be explored.
We are also planning to improve the infection model to find effective strategies
for the attacker to launch automatic infection while limiting the detection
factor.
\bibliographystyle{plain}
\bibliography{tormalware.bib}
\end{document}